\documentclass[usenatbib]{mn2e}
\usepackage{graphicx}
\usepackage{rotating}
\usepackage{blindtext}
\usepackage{longtable}
\usepackage{lscape}
\usepackage{url}
\usepackage{amstext}
\usepackage{color}

\title[NS Energy Loss Mechanisms]
  {On the Diversity of Compact Objects within Supernova Remnants II: Energy Loss Mechanisms}
\author[Adam Rogers and Samar Safi-Harb]
{Adam Rogers and Samar Safi-Harb\thanks{Canada Research Chair} \\ Department of Physics and Astronomy, University of Manitoba, Winnipeg, Manitoba, Canada, R3T 2N2}

\date{Accepted 2016 October 14. Submitted 2016 July 23.}
\pagerange{\pageref{firstpage}--\pageref{lastpage}} \pubyear{2016}

\begin{document}
\label{firstpage}
\maketitle

\begin{abstract}
Energy losses from isolated neutron stars are commonly attributed to the emission of electromagnetic radiation from a rotating point-like magnetic dipole in vacuum. This emission mechanism predicts a braking index $n=3$, which is not observed in highly magnetized neutron stars. Despite this fact, the assumptions of a dipole field and rapid early rotation are often assumed a priori, typically causing a discrepancy between the characteristic age and the associated supernova remnant (SNR) age. We focus on neutron stars with `anomalous' magnetic fields that have established SNR associations and known ages. Anomalous X-ray pulsars (AXPs) and soft gamma repeaters (SGRs) are usually described in terms of the magnetar model, which posits a large magnetic field established by dynamo action. The high magnetic field pulsars (HBPs) have extremely large magnetic fields just above QED scale (but below that of the AXPs and SGRs), and central compact objects (CCOs) may have buried fields that will emerge in the future as nascent magnetars. In the first part of this series we examined magnetic field growth as a method of uniting the CCOs with HBPs and X-ray dim isolated neutron stars (XDINSs) through evolution. In this work we constrain the characteristic age of these neutron stars using the related SNR age for a variety of energy loss mechanisms and allowing for arbitrary initial spin periods. In addition to the SNR age, we also make use of the observed braking indices and X-ray luminosities to constrain the models.
\end{abstract}


\section{Introduction}
\label{sec:intro}

In previous work \citep[][hereafter RSH16]{rsGrowth} we studied a parametrized phenomenological model for describing magnetic field growth in neutron star (NS) evolution. Assuming the external dipole field is buried by an intense process of fall-back accretion after the formation of the NS \citep{BGrowthRef1, BGrowthRef2}, the slow diffusion from the stellar surface exerts a time-dependent torque on the NS \citep{BGrowthRef3}. This process may provide an explanation for the observed braking indices of young pulsars with $n \neq 3$ \citep{nReview, bigN}, in contrast to the prediction of a rotating magnetic dipole in vacuum \citep[$n=3$,][]{ostrikerGunn69}. Magnetic field growth provides an evolutionary link between apparently disparate classes of NS, such as the extremely weak field\protect{\footnote{In this work we make use of the dipole magnetic field at the stellar equator and poles, denoted with subscripts \textit{eq} and \textit{p} respectively, and defined as $B_{eq}=3.2 \times 10^{19}\sqrt{ P \dot{P} }$~(G) and $B_{p}=2 B_{eq}$. The field is inferred from the observed period, $P$(s), and period derivative, $\dot{P}$ (s s$^{-1}$).}} subset of NSs known as central compact objects (CCOs), high magnetic field pulsars (HBPs) and the X-ray dim isolated NSs (XDINSs). On the other hand, magnetic field decay has been invoked to describe the evolution of the anomalous X-ray pulsars (AXPs) and soft gamma repeaters (SGRs). These objects are conventionally described by the magnetar model, which posits the development of large magnetic fields via dynamo action from rapid rotation early in the life of the NS, necessary to support the core of the massive progenitor from collapse \citep[][]{dunThom92, Akiyama03, thom05}. However, the distinction between the apparently rotation-powered HBPs, with dipole fields just above the quantum critical limit, $B_\text{QED}=4.4 \times 10^{13}$ G, and the magnetically powered AXPs and SGRs ($B\geq B_\text{QED}$) is significantly blurred after the HBPs were observed displaying magnetar-like activity \citep{gavriil08, harshaSamar2008, J1119burst}, and radio emission was observed from magnetars \citep{1E1547}. Moreover, the SNRs associated with the AXPs and SGRs show evidence of `typical' explosion energies ($\leq10^{51}$ erg) and not super-energetic as one would expect from a rapidly rotating proto-NS \citep{noEnergeticSNR, harsha1, samarHarshaMWN, 1841Age}. These complications are difficult to reconcile with the standard magnetar picture.

The SNR association can also impose a profound constraint on the NS energy loss, since the NS characteristic age $\tau$ must agree with the age of the associated SNR $\tau_{\text{SNR}\pm}$ (denoting the lower SNR age with a minus and the upper limit with a plus). Generally the errors on SNR ages are large, and depend sensitively on both the distance to the remnant and its expansion velocity or assumed phase of evolution. Despite these large error bars, poor agreement is generally still found with the NS characteristic age, in some cases by orders of magnitude. The CCOs present extreme examples of this dichotomy. These objects appear to be extremely old NSs, but they are intrinsically young, being associated with young SNRs.

The discrepancy between neutron star and SNR age has been recently studied by several authors \citep[e.g.,][RSH16]{charAge2012, gao2014, nakano15}. \citet{charAge2012} calculate the initial spin period $P_0$ as a function of SNR age and constant braking index for a set of $30$ radio pulsars associated with SNRs. \citet{gao2014} and \citet{gao2015} address the magnetar-SNR associations, however they include uncertain associations, SNRs with uncertain or unknown ages, and assume that $P_0 \ll P$. \citet{nakano15} explain the large discrepancy between the age of SNR CTB~109 and its AXP 1E~2259+586 by magnetic field decay, however they also assume a very small initial spin period. In this work we explore the relationship between NS and SNR ages for a selection of energy-loss mechanisms with arbitrary initial period $P_0$, making use of the measured NS properties (period $P$, period derivative $\dot{P}$, braking index $n$ and X-ray luminosity $L_x$). The properties of the secure NS-SNR pairs that are considered in this work are summarized in Table \ref{tableSys}. The SNR ages and distance estimates were collected from the University of Manitoba's high-energy catalogue of SNRs, SNRcat\footnote{http://www.physics.umanitoba.ca/snr/SNRcat} \citep{SNRCat}, while the McGill pulsar catalog\footnote{http://www.physics.mcgill.ca/~pulsar/magnetar/main.html} provides the properties of the AXPs and SGRs. In Section \ref{sec:mech}, we discuss the general expression relating a constant braking index and initial spin period for power-law torques. Section \ref{sec:B(t)} introduces time-dependent magnetic fields and describes magnetic field evolution. We discuss braking by the emission of relativistic winds in Section \ref{sec:wind}. We provide further discussion of our results in Section \ref{sec:discussion} and summarize our conclusions in Section \ref{conclusions}. Justification for excluding some of the NS-SNR pairs is discussed in Appendix \ref{appA}.

\section{Braking Mechanisms}
\label{sec:mech}

The spin-down energy loss from a rotating NS is $\dot{E}=-I\Omega\dot{\Omega}$, with $\Omega=2\pi/P$. The braking index determines the torque acting on the NS and is defined as
\begin{equation}
n=\frac{\Omega \ddot{\Omega}}{\dot{\Omega}^2}
\end{equation}
which allows a powerful probe of the physics involved in pulsar spin-down if the period and derivatives can be measured accurately. We summarize the NS energy loss mechanisms below.

\subsection{Power-Law Torques}
\label{PLT}

\begin{figure}
\centerline{\includegraphics[scale=0.7, bb= 219 164 374 637, clip=true]{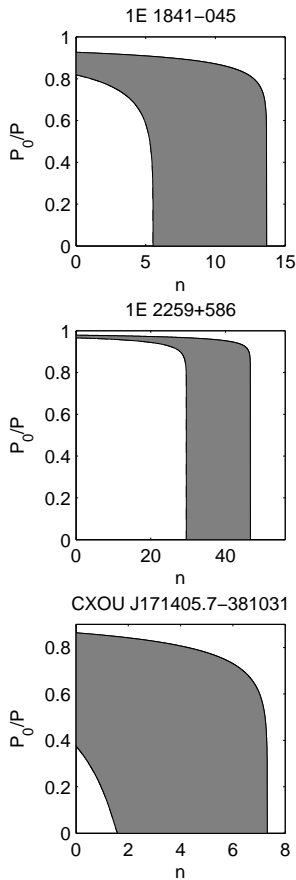}}
\caption{Power-law age constraints for the AXPs: 1E~1841--045, 1E~2259+586, and CXOU J171405.7--381031. Gray areas indicate regions where the pulsars' ages agree with the SNR age estimates (see table \ref{tableSys}). We show that large ratios of initial and observed spin periods are needed for a large range of braking indices using a power-law spin-down torque.}
\label{fig1A}
\end{figure}

\begin{figure}
\centerline{\includegraphics[scale=0.7, bb= 134 251 459 540, clip=true]{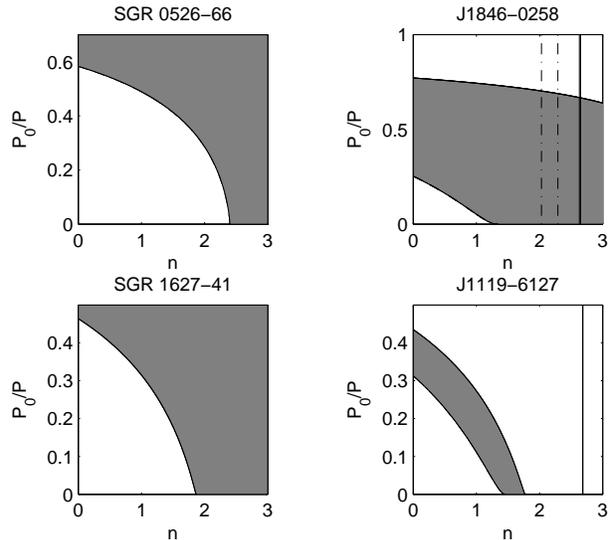}}
\caption{Power-law age constraints. Left column: SGRs 0526--66 (top) and 1627--41 (bottom). Right column: HBPs J1846--0258 (top) and J1119--6127 (bottom). See Table \ref{tableSys} and text for details. The vertical lines included in the plots of the HBPs mark the measured braking indices including uncertainties. The pre-outburst and post-outburst braking indices of HBP J1846--0256 are shown in solid and dash-dotted lines, respectively.}
\label{fig1B}
\end{figure}

The spin-down torque acting on a NS is often expressed as a power-law in terms of a constant braking index; the emission of magnetodipole radiation from a point source in vacuum has $n=3$ and that of quadrupole radiation (both gravitational and electromagnetic) has $n=5$, assuming constant fields. For constant $n$, we express the initial spin period $P_0$ in terms of the braking index \citep{glendenningBook}, such that
\begin{equation}
P_0=P\left[ 1-(n-1)\tau \frac{\dot{P}}{P} \right]^{\frac{1}{(n-1)}}.
\label{P0}
\end{equation}
Equating the PSR age, $\tau$, with the independently measured SNR age gives the initial period $P_0$ as a function of $n$ ($P$ and $\dot{P}$ are known from timing observations). The results are shown in figures \ref{fig1A} and \ref{fig1B}. The shaded areas in these plots are bounded by the SNR ages, $\tau_{SNR \pm}$, which determine the spread in the initial spin period $P_0/P$ and the braking index $n$. These plots show that small initial spin periods are excluded for several of the sources despite the typical a priori assumption that $P_0 \ll P$. AXP 1E~2259+586 is a particularly severe case, as the age of CTB~109 requires a braking index $n>30$ for small $P_0$. The inclusion of a quadrupole magnetic field component ($n=5$) does not alleviate this problem, particularly since the morphology of the X-ray light curves are well fit by polar hot spot models indicative of a global dipole field \citep{pulse1, pulse2, pulse3, pulse4}. Thus if the magnetar model is a realistic description, the solution of the age problem depends on variable braking indices of NSs. Time-dependent fields, described in Section \ref{sec:B(t)}, offer a larger range of dynamical behavior and accommodate the observed variety of NS properties including the corresponding SNR ages.

The CCOs are not plotted in these figures as they require large initial spin periods ($P_0>0.99 P$) for braking index in the range $0<n<50$. These large initial spin periods are insufficient to generate a strong dynamo and imply a weak magnetic field. This scenario forms the basis of the proposed ``anti-magnetar'' model of CCOs, which posits an extremely weak field orders of magnitude lower than the magnetar model and X-ray luminosity due to thermal emission from residual cooling \citep{firstAntiMagnetar, gotthelf13}, though the CCO 1E~16348-5055, associated with the SNR RCW~103, has been observed to display magnetar-like bursting behaviour \citep{RCW103burst, RCW103burst2}. The constant braking index assumption with the observed SNR age constraints naturally leads to this view since $P_0 \approx P$ is naturally recovered. However, when more complex evolutionary scenarios are considered this condition can be alleviated, leading to alternatives to the high $P_0/P$ ratios of the CCOs. Note that the SGRs in figure \ref{fig1B} have no lower SNR age estimates and so are only gently constrained. 

Systems with a measured braking index provide interesting examples with further constraints. The HBP J1119--614 associated with the SNR G292.2--0.5 has $n=2.684\pm 0.002$ \citep{welt11}. However, in order for the pulsar to have an age that is consistent with the SNR, we expect the braking index to be much lower than this value. As shown in figure \ref{fig1B}, we find that the braking index is constrained by the SNR's age to be $1.45 < n < 1.8$ provided $P_0 \ll P$, consistent with an earlier estimate $n<2$ \citep{harsha1}. We use the period, period derivative and braking index determined by \citet{welt11}, however we note that \citet{n1119_2015} have a measured value $n=2.677$, which does not follow from the spin parameters in that work due to the details of the analysis performed. The small difference between the two braking indices will not significantly affect the results. The HBP J1846--0258 demonstrates that the braking index varied over time. Measurements of the spin period and derivatives were made before and after a magnetar-like outburst detected in 2006 \citep[][]{livingstone06, gavriil08, harshaSamar2008}, and show changes in the pre- and post-outburst braking index. We include both in figure \ref{fig1B}, with the measurements made from 2000-2003 \citep[$n=2.64\pm 0.01$;][]{livingstone06} and 2008-2010 \citep[$n=2.16 \pm 0.13$;][]{livingstone11}. The braking index has remained at comparatively low values \citep{archibald15}. The physics leading to the constancy of the X-ray luminosity and pulse profile from the pre- to post-outburst phase is difficult to explain in most conventional models. Whether the braking index will eventually return to the pre-outburst value remains an open question. Monitoring this object is crucial to address this question.

\subsection{Magnetic Field Evolution}
\label{sec:B(t)}

A popular explanation for magnetar X-ray luminosity in excess of the spin-down energy loss is the release of energy stored in the magnetic field. For NSs containing $npe$ matter and neglecting superfluid processes, magnetic field decay channels were first studied by \citet{goldreichReisenegger1992}, and depend on a variety of effects including Ohmic dissipation, ambipolar diffusion and Hall drift  \citep{TD96, BdecaySurf, thompson2002, beloborodov2009, hall2014}. Let us compare the timescales over which these effects are expected to be significant. In terms of the average conductivity of NS matter, $\sigma_{0}=5 \times 10^{26}$ s$^{-1}$ and a crust thickness of $L=1$ km \citep{hallRef1}, we find the timescale for Ohmic dissipation,
\begin{equation}
 \tau_{Ohm}=\frac{4\pi \sigma_0 L^2}{c^2},
\end{equation}
which gives a timescale on the order of $\tau_{Ohm} \sim 10^6$ kyr. However, this is a simplification due to the dependence of $\sigma_0$ on electron density and temperature. Changes in these quantities can cause variations in $\sigma_0$ over orders of magnitude \citep{hallRef2}, and for low-field NSs with ages $\sim 10$ to $100$ kyr, Ohmic dissipation is expected to dominate the magnetic field evolution.
In contrast, for high field NSs, the timescale for Hall drift in terms of the electron number density $n_{e}$ and charge $e$ is
\begin{equation}
 \tau_{Hall}=\frac{4 \pi n_{e} e L^2}{c B},
 \label{hallTau}
\end{equation}
which gives $\frac{10}{B_{14}}$ kyr $<\tau_{Hall}< \frac{1000}{B_{14}}$ kyr with $B_{14}$ the magnetic field in units of $10^{14}$ G, and electron densities between $10^{35}$ and $10^{37}$ cm$^{-3}$. Thus, a magnetar-scale field of $10^{15}$ G can evolve on the order of $1-100$ kyr. Therefore the Hall effect provides the dominant mechanism for the field evolution of highly magnetized, young NSs on timescales comparable with the observed ages of SNRs \citep{hallRef1, hallRef2}.

A parameterized phenomenological model of time-dependent magnetic field decay was developed by \citet{colpi2000} and expanded upon by \citet{dall'osso}. The model posits a decaying dipole field such that
\begin{equation}
P\dot{P}=bB_{\text{eq}}^{2}(t)
\label{PdotP}
\end{equation}
with $b$ a constant, the period and derivative now carry an implicit time-dependence that comes from the time evolution of the magnetic field $B_\text{eq}$. We assume a point-dipole in vacuum that is an orthogonal rotator such that $\chi(\theta)=\sin^2\theta=1$ with $\theta$ constant, but magnetospheric effects can affect this assumption \citep{magneto2}.

In general, we describe both field growth and decay using the same basic framework (RSH16). Suppose the magnetic field evolves in time as
\begin{equation}
B_{\text{eq}}(t)=B_{\text{j}} f_{\text{j}}(t)
\label{Bt}
\end{equation}
where $B_{\text{j}}$ is a constant equatorial reference field, either the initial field in decay models \citep[labelled as $j=D$,][]{igoshev1, igoshev2} or the final field strength in growth models (labelled as $j=G$, RSH16). The function $f_{\text{j}}(t)$ carries the time-dependence. The characteristic age is defined here as
\begin{equation}
\tau=\frac{P}{2\dot{P}}=\frac{P_{\text{0}}^2}{2bB_{\text{j}}^2f_{\text{j}}^2}+\frac{F_{\text{j}}^2}{f_{\text{j}}^2}.
\label{tau}
\end{equation}
where
\begin{equation}
F_{\text{j}}^2=\int_0^t f_j^2(t')dt'.
\label{Fj2}
\end{equation}
When evolution is driven by a changing field, the model time $t$ describes the true age of the NS, and the characteristic age $\tau_{\text{PSR}}$ gives the apparent observable age of the NS. Thus, we match $\tau_{\text{PSR}}$ to the observed values given in table \ref{tableSys} for a true age $\tau_{\text{PSR}-} < t < \tau_{\text{PSR}+}$. The braking index is then time-dependent,
\begin{equation}
n= 3 - 4 \tau \frac{\dot{f_{\text{j}}}}{f_{\text{j}}}.
\label{n}
\end{equation}
Thus, field decay gives $n>3$ since $\dot{f_{\text{D}}}<0$, and field growth gives $n<3$ since $\dot{f_{\text{G}}}>0$ . This braking index formula is similar to the effect of including superfluidity in the NS interior via a varying moment of inertia due to an increasing fraction of the NS core converted to the superfluid state over time through neutrino cooling \citep[][for example]{hoAn2012}. Then the braking index has the same form as equation \ref{n}, with the substitution $f_{\text{j}} \rightarrow I$.

First let us consider field decay, which enhances the apparent age of an NS, and thus acts on objects with $\tau_{\text{PSR}} > \tau_{\text{SNR}}$. We adopt the following parametric form for the decaying field \citep{colpi2000, dall'osso}
\begin{equation}
B(t)=B_{\text{D}} \left\{
\begin{array}{ll}
\left( 1+ \alpha \frac{t} {\tau_{\text{D,0}}} \right)^{-\frac{1}{\alpha}} , & \alpha\neq0,2 \\
\exp\left( -\frac{t}{\tau_{\text{D,0}}} \right), & \alpha=0 \\
\end{array}\right.
\label{BD}
\end{equation}
where the dynamic decay timescale $\tau_{\text{D,t}}=\left[ a B(t)^{\alpha} \right]^{-1}$ is given in terms of the field at time $t$, and a normalization constant $a$. This timescale changes from its initial value $\tau_{\text{D},0}$ as the field decays.

\begin{figure}
\centerline{\includegraphics[scale=0.65, bb= 120 160 467 638, clip=true]{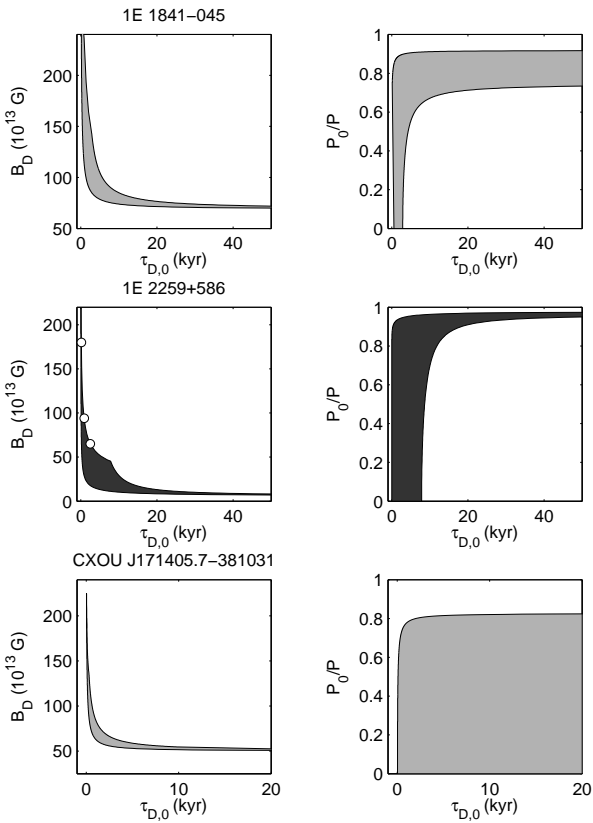}}
\caption{Field decay age constraints for the AXPs 1E~1841--045 (top row), 1E~2259+586 (middle row) and CXOU~J171405.7-- 381031 (bottom row). In the left column we plot the initial field $B_\text{D}$ in units of $10^{13}$ G vs. the initial decay timescale $\tau_{\text{D},0}$ in kyr, and in the right column we plot the initial period fraction $P_0/P$ against the timescale. These parameter space regions are calculated for a variety of $\alpha$ between $0$ and $2$. Solutions that satisfy the SNR age constraints are shown as the shaded regions. Light gray indicates decay models that are capable of accommodating the observed X-ray luminosity. The solutions for 1E 2259+586 reported by \citet{nakano15} are plotted as white disks on the $B_\text{D}-\tau_\text{D,0}$ plot. These solutions have $P_{\text{0}}=3$ ms, so occupy the region near the $\tau_{\text{D},0}$ axis and are not shown on the associated right-hand panel.}
\label{bDecayAXP}
\end{figure}

\begin{figure}
\centerline{\includegraphics[scale=0.80, bb= 148 248 440 545, clip=true]{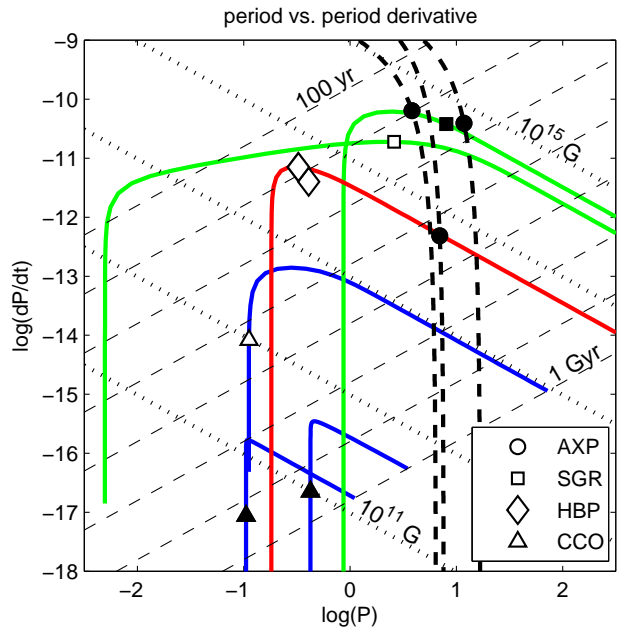}}
\caption{Period vs. period derivative. Lines of constant characteristic age are plotted as dashed lines ranging from $100$ years (upper left) to $1$ Gyr (lower right), increasing by factors of $10$. Lines of constant magnetic field are plotted as dashed lines from $10^{11}$ G (lower right) to $10^{15}$ G (upper left), increasing by factors of $10$.
AXPs are plotted as circles and have dashed evolutionary trajectories given by field decay. The SGRs are squares and have green field growth trajectories. HBPs are diamonds and the J1846--0258 field growth path is plotted in red. Note that we do not provide a trajectory for HBP~J1119--6127 (see text for details). Finally, the CCOs are triangles with field growth paths shown in blue. Sources with $L_x<\dot{E}$ are hollow and sources with $L_x>\dot{E}$ are filled. Both axes are plotted on logarithmic scales.}
\label{figPPdotBEvolution}
\end{figure}

To compare the model and observable quantities, we restate the field decay model in terms of the observed period and period derivative labelling the dipole field at time $t$ as $B_\text{t}$ and characteristic age $\tau(t)=\tau_\text{PSR}$. Using the observed field in equation \ref{BD}, we solve for the model time
\begin{equation}
t=\left\{
\begin{array}{ll}
\frac{\tau_\text{D,0}}{\alpha} \left[ \left( \frac{B_\text{0}}{B_\text{t}} \right)^{\alpha} -1 \right], & \alpha\neq 0 \\
\tau_\text{D,0} \ln\left( \frac{B_\text{0}}{B_\text{t} } \right), & \alpha = 0
\end{array}
\right.
\end{equation}
provided that $B_\text{D} >  B_\text{t}$. Then, equation \ref{tau} allows us to write the initial period as a function of the characteristic age and the free parameters
\begin{equation}
P_\text{0} = \sqrt{ 2bB_\text{eq}^2\left( \tau_\text{PSR} f_D^2 - F_D^2 \right) }
\label{dynamicP0}
\end{equation}
which requires
\begin{equation}
y=\tau_\text{PSR} f_D^2 - F_D^2 \geq 0.
\end{equation}
Thus, using the observed period and period derivative simplifies the problem, since we calculate $P_\text{0}$ in terms of the observed quantities and the three remaining parameters $\alpha$, $\tau_\text{D,0}$ and $B_\text{D}$. To find the set of solutions that obey the age constraint, we fix $\alpha$ at a given value and calculate $t$ for each ($\tau_\text{D,0}$, $B_\text{D}$) pair between the physically relevant lower and upper bounds. We limit the initial field $B_\text{D}$ between the maximum observed magnetar field of SGR~1806--20, with $B_\text{max}=2.40\times 10^{15}$ G \citep{maxB} and the observed field $B_\text{t}$. We let the decay timescale vary from $1$ to $100$ kyr, the limit of SNR observability. The model age $t$ can then be contoured at the values $t=\tau_\text{SNR-}$ and $t=\tau_\text{SNR+}$ to give the set of solutions that satisfy the age constraint. For each of the pairs of parameters, $y$ can also be calculated and contoured at $y=0$ to give the boundary between physical and unphysical solutions. The regions of the parameter space which simultaneously satisfy the age constraint with $y>0$ are valid solutions. This process is repeated for $\alpha$ from $0$ to $2$ to thoroughly explore the parameter space. In the left column we plot the inital field $B_\text{D}$ as a function of the initial decay timescale, and in the right column we show the corresponding ratio of initial to observed period $P_\text{0}/P$ against the decay timescale. The luminosity produced by the decay of a dipole field over the decay timescale is \citep{dall'osso}
\begin{equation}
L_{\text{D,t}}=-\frac{ R^3 B_{\text{D,t}}^2 }{ 3 \tau_{\text{D,t}} };
\label{LD}
\end{equation}
where the currently observable luminosity is then $L_{\text{D,t}}$, using $B_{\text{D,t}}$ and $\tau_{\text{D,t}}=\tau_\text{D,0}+\alpha t$. For each of the AXPs, the results are shown in Figure \ref{bDecayAXP}. The solutions given by \citet{nakano15} are marked for AXP~1E~2259+586 as white disks. Regions in Figure \ref{bDecayAXP} that are colored light gray have observed luminosity $L_\text{x} \leq L_\text{D,t}$ and otherwise appear dark gray. We plot the observed period and period derivatives of the systems from table \ref{tableSys} in figure \ref{figPPdotBEvolution}. In this figure we also include lines of constant $\tau_{PSR}=P/2\dot{P}$ (thin dashed lines), and constant dipole field (thin dotted lines). NSs that have X-ray luminosity greater than spin down luminosity are marked by filled symbols. We show example evolutionary trajectories for the AXPs as thick black dashed curves.

\begin{figure}
\centerline{\includegraphics[scale=0.80, bb= 157 246 446 545, clip=true]{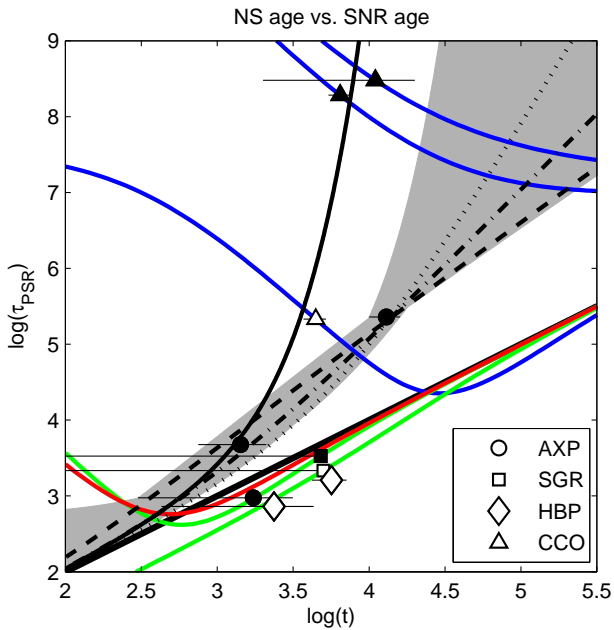}}
\caption{Pulsar age vs. SNR age. The horizontal lines show the spread in SNR age estimates, ranging from $\tau_{SNR-}$ to $\tau_{SNR+}$ (see Table \ref{tableSys}). If one of these limits is not available, we plot the error bar to the corresponding axis limit. AXPs are plotted as circles, SGRs are squares, HBPs are diamonds and CCOs triangles. Sources with  $L_x<\dot{E}$  are not filled, and sources with $L_x>\dot{E}$ are filled. Each of the axes are plotted on a logarithmic scale. The diagonal black line corresponds to $\tau=t$. The light gray region is our joint fit region for field decay. The solutions found by Nakano are plotted as heavy dashed ($\alpha=0.6$), dash-dotted ($\alpha=1.0$) and dotted ($\alpha=1.4$) lines. The heavy solid black line is our best joint-fit solution found for the CCOs. The details of these solutions are given in Table \ref{tableBEvol}. Note that we do not provide a trajectory for HBP~J1119--6127.}
\label{figTAge}
\end{figure}

We also performed a joint fit for the AXPs and fit the observed spread of $\tau_\text{PSR}$ and $\tau_{\text{SNR}\pm}$ in terms of 1E~2259+586 since it has the largest characteristic age among the AXPs. In this view the age discrepancy is an evolutionary effect. The relationship between $\tau_{\text{PSR}}$ and $\tau_{\text{SNR}\pm}$ for all systems in table \ref{tableSys} is plotted in figure \ref{figTAge}. This figure also includes the $\tau_{\text{PSR}}=\tau_{\text{SNR}}$ line, shown in heavy black. Since we consider general initial spin periods in our models, our parameter sets include the solutions found by \citet{nakano15}, who assumed a constant $P_0 \ll P$. The \citet{nakano15} solutions are marked as dashed black lines, and have $\alpha$ values $0.60$, $1.00$ and $1.40$, where \citet{colpi2000} favour $\alpha=1$. Our joint fits occupy the light gray region in Figure \ref{figTAge}. Following this approach we also consider a joint fit to the CCOs in terms of CXOU J185238.6+004020. The only viable field decay mode for the CCOs requires exponential decay since the characteristic ages are so long for these objects. Since exponential decay is disfavoured among magnetic field decay modes \citep{dall'osso, nakano15}, this result suggests that field decay is unlikely to properly describe the CCO evolution, which we suggest is more neatly accommodated by field growth (RSH16). The parameters that describe the joint fit are given in table \ref{tableBEvol}.

\setcounter{table}{1}
\begin{table*}
\begin{center}
\begin{tabular}{|l|l|l|l|l|l|l|}
\multicolumn{7}{c}{ } \\ \hline
\multicolumn{7}{c}{Solutions Shown in Figure \ref{figTAge}} \\
\multicolumn{7}{c}{Joint Field Decay} \\ \hline

Object & $\alpha$ & -- & $\tau_{\text{D},0}$ (yr) & $B_{\text{D,0}}$ ($\times 10^{13}$ G) & $P_0$ (ms) & --\\ \hline
AXP & $0.60$     & & $2500$ & $65.00$ & $3.0$ & \\
AXP & $1.00$     & & $920$ & $94.00$ & $3.0$ & \\
AXP & $1.40$     & & $160$ & $180.00$ &$3.0$ & \\ \hline
CCO & $0.00$  &  & $1180$ & $3.48$ &  $421.00$ & \\ \hline

\multicolumn{7}{c}{Individual Field Growth} \\
Object & $\alpha$ & $\epsilon$ ($\times 10^{-5}$)& $\tau_{\text{G}}$ (yr) & $B_{\text{G}}$ ($\times 10^{13}$ G) & $P_0$ (ms) & $t$ (kyr)\\ \hline
SGR 0526--66 & $0.155$ & $0.007$ & $789.730$ & $57.581$ & $0.868$ & $4.800$ \\
SGR 1627--41 & $0.319$ & $2.023$ & $3700.000$ & $41.763$ & $0.005$ & $5.000$ \\ \hline
HBP J1846--0258 A & $0.000$ & $6.372$ & $252.686$ & $5.178$ & $0.182$ & $0.900$ \\
HBP J1846--0258 B & $0.341$ & $5.487$ & $223.731$ & $6.014$ & $0.182$ & $0.900$ \\ \hline
CCO RX J0822.0--4300 & $0.202$ & $818.016$ & $33550.160$ & $0.909$ & $0.112$ & $4.450$ \\
CCO 1E 1207.4--5200 & $0.700$ & $100$ & $98245.295$ & $0.045$ & $0.423$ & $11.000$ \\
CCO CXOU J185238.6+004020 & $0.468$ &  $0.017$ &  $12483.740$ & $0.014$ & $0.105$ & $6.450$ \\ \hline

\end{tabular}
\caption{Field evolution solutions plotted in figure \ref{figTAge}. AXP solutions from \citet{nakano15}.}
\label{tableBEvol}
\end{center}
\end{table*}

Next let us consider field growth. The burial of a magnetic field by fall-back accretion and the resulting consequences for NS evolution was first studied by \citet{BGrowthRef1, BGrowthRef2} and \citet{BGrowthRef3}. We use the phenomenological growth model from RSH16 to demonstrate the effect of magnetic field emergence. In place of equation \ref{BD}, consider a growing field
\begin{equation}
B(t) = B_{\text{G}} \left\{
\begin{array}{ll}
1+\epsilon-\left( 1+\alpha \frac{t}{\tau_{\text{G}}} \right)^{\frac{\alpha-1}{\alpha}}, & 0 < \alpha < 1 \\
1+\epsilon-\exp\left(-\frac{t}{\tau_{\text{G}}}\right), & \alpha=0
\end{array}\right.
\label{BG}
\end{equation}
where $B_\text{G}$ is the magnetic field after field growth, the growth timescale is the constant $\tau_{G}$, $\alpha$ is the growth index and $\epsilon$ an additive constant which determines the initial field in terms of the final asymptotic field $B_{\text{G},0}=\epsilon B_{\text{G}}$ \citep{nb15}. In this framework the only formal changes between the decay and growth models are the details of the field evolution function $f_{\text{G}}$ and its integral $F_{\text{G}}^2$, with analytical expressions given in RSH16. The model time in terms of the free parameters ($\alpha$, $\epsilon$, $\tau_\text{G}$, $B_\text{G}$) and the observed dipole field at the present time, $B_\text{t}$, is
\begin{equation}
t=\left\{
\begin{array}{ll}
\frac{\tau_\text{G}}{\alpha} \left[ \left( 1+\epsilon-\frac{B_\text{t}}{B_\text{G}} \right)^{\frac{\alpha}{\alpha-1}} -1 \right], & \alpha\neq 0 \\
-\tau_\text{G} \ln\left( 1+\epsilon-\frac{B_\text{t}}{B_\text{G} } \right), & \alpha = 0
\end{array}
\right.
\end{equation}
with $P_\text{0}$ given by equation \ref{dynamicP0} for $f_\text{G}$ and $F_\text{G}^2$. The constraints on NSs with measured braking indices are discussed at length in RSH16. We find solutions for systems with $\tau_{\text{PSR}} < \tau_{\text{SNR}}$ and use the constraint from the measured braking indices when available. We display example evolutionary trajectories for the HBP J1846--0258 (post-outburst) and assuming the upper SNR age $\tau_{\text{SNR}\pm}$ for the SGRs in figure \ref{figPPdotBEvolution} and figure \ref{figTAge}.

For the HBP J1119--6127, we were unable to find a braking index in the observed range given the age limits of the associated SNR, G292.2--0.5 \citep{harsha1}, and found a value $n \approx 1.2$ (RSH16). This differs significantly from the measured braking index, $n=2.684 \pm 0.002$. As an alternative, we have also fit J1119--6127 by searching for solutions that satisfy the braking index constraint and determining the resulting age. This approach gives a maximum model age $t_{\text{SNR}} \approx 1.76$ kyr for $n$ in the observed range, well short of the $4.2$ kyr lower SNR age limit for G292.2--0.5. Due to these discrepancies with observation we do not show a trajectory for HBP~J1119--6127 in Figures \ref{figPPdotBEvolution} and \ref{figTAge}.

\subsection{Braking by relativistic wind}
\label{sec:wind}

In general, the Hall timescale for magnetic field evolution depends on the strength of the magnetic field, as seen in equation \ref{hallTau}. For young NSs with fields below $10^{14}$ G, this timescale may be longer than the observed SNR age. Therefore magnetic field growth does not have a dramatic effect upon these young NSs. However, many of these systems are observed to have a braking index $n<3$ \citep{nReview, bigN}. A possible explanation for these low braking indices may be through the emission of a relativistic particle wind \citep{TB98, harding99, windbraking2013}. However, the conclusive detection of wind nebulae around magnetars in particular is challenging due to the presence of dust-scattering haloes that accompany these X-ray bright, heavily absorbed objects \citep{dustScattering, samarMWN}. Only a handful of such nebulae have been proposed to be associated with highly magnetized NSs. For example, a wind nebula has been proposed to surround the magnetar Swift~J1834.9--0846 in W41 \citep{younes16}, and the luminosity of a particle wind was estimated for SGR~1806--20 based on the X-ray and radio observations of the wind-powered nebula G10.0--0.3 \citep{TD96, marsden1999, gaenslerWind}. In the pulsar wind model, relativistic particles load the magnetosphere with charge and distort the dipole field at large scales outside of the light cylinder. Besides affecting the NS spin-down, the emission of a relativistic wind can also offer an explanation for the significant timing noise that generally affects magnetar observations \citep{timingNoise}. The HBPs J1119--6127 and J1846--0258 are clearly associated with pulsar wind nebulae \citep{samarHarsha1119, gavriil08, harshaSamar2008, ng1846, samarMWN}, suggesting that particle wind emission should play an important role in their evolution. We also expect that AXP and SGR evolution may be affected by wind emission due to candidate wind nebulae, but do not consider these models for the CCOs which do not show any evidence of PWN. However, we note that braking exclusively due to a steady particle wind produces a torque with a braking index $n=1$, too low for the NSs with secure SNR associations in table \ref{tableSys}.

A more realistic scenario is energy loss from a time-varying particle wind and magnetodipole energy loss simultaneously \citep{TD96}. We include both effects using a duty-cycle in which the neutron star undergoes periods where wind-braking is dominant, and periods where the wind is switched off \citep{harding99, windbraking2013, windbraking2014}. This transition allows for braking indices in the range $1 < n < 3$ which agrees with the measured braking indices of the HBPs. The duty cycle $D_p$ gives the fraction of energy lost as wind with instantaneous kinetic luminosity $L_p$. Since the particle and X-ray luminosity in magnetars are related to magnetic processes we expect $L_x \approx L_p$. The average energy loss is
\begin{equation}
\dot{E}_\text{wd}= \dot{E}_\text{d}\eta_\text{wd}
\label{EdotWindDipole}
\end{equation}
with
\begin{equation}
\eta_\text{wd}=(1-D_\text{p})+\left(\frac{L_\text{p}}{\dot{E}_\text{d}}\right)^{ \frac{1}{2} } D_\text{p},
\label{etaWD}
\end{equation}
Following \citet{harding99}, the total polar field is
\begin{equation}
B_\text{p}=-\frac{\sqrt{6c^3}}{8 \pi^2} \frac{L_\text{p}^{1/2}D_\text{p} P^2 }{(1-D_\text{p}) R^3}F(P,\dot{P})
\label{BwindDipole}
\end{equation}
in terms of the function
\begin{equation}
F(P,\dot{P}) = 1-\left[ 1+\frac{4\dot{E}(1-D_\text{p})}{L_\text{p} D_\text{p}^2} \right]^{1/2}
\end{equation}
with the average rotational energy loss $\dot{E}=4\pi^2I\frac{<\dot{P}>}{P^3}$ for the mean period derivative, which we take to be the observed value. Integrating Equation \ref{EdotWindDipole} from $P_\text{0}$ to $P$ gives the general result for the characteristic age due to both wind braking and dipolar emission,
\begin{equation}
\tau_\text{dw} = -\frac{ 4 \pi^2 I (1-D_\text{p}) }{ L_\text{p} D_\text{p}^2 P^2 } \frac{1}{F(P,\dot{P})} \ln \left(    \frac{1-\frac{2}{F(P,\dot{P})} }{1-\frac{2 P_\text{0}^2}{P^2 F(P, \dot{P})}}   \right).
\label{tauWindDipole}
\end{equation}
Equations \ref{BwindDipole} and \ref{tauWindDipole} reduce to the dipole model when $L_\text{p}D_\text{p}^2 \ll 4\dot{E}(1-D_\text{p})$ and $D_\text{p} \rightarrow 0$. The steady wind model is recovered when $L_\text{p} D_\text{p}^2 \gg 4\dot{E}(1-D_\text{p})$. The braking index for the magnetar wind model is
\begin{equation}
n = 3 + B_\text{p} R^3 D_\text{p}\frac{\Omega}{I\dot{\Omega}} \left( 2-\frac{\Omega}{2\dot{\Omega}}\frac{\dot{L_\text{p}}}{L_\text{p}} \right)\sqrt{\frac{L_\text{p}}{6c^3}}.
\label{nWind}
\end{equation}
For completeness, we include the contribution from a time dependent luminosity $\dot{L}_{\text{p}}$. For a constant luminosity and $D_{\text{p}}=1/2$, we recover the well-known braking index formula for combined wind and magnetodipole spin-down \citep[][]{livingstone11, gao2014, archibald15}. This emission model requires three free parameters: the wind luminosity, $L_\text{p}$, the duty cycle parameter $D_\text{p}$, and the initial spin period $P_\text{0}$. We allow for $0 < P_\text{0}/P < 1$, and find the set of parameters which give the pulsar characteristic age bounded by the SNR age limits $\tau_{\text{SNR}-} \leq \tau_{\text{DW}} \leq \tau_{\text{SNR}+}$ using the contouring approach described in section \ref{sec:B(t)}. We numerically integrate $\dot{\Omega}$ from equation \ref{EdotWindDipole} to avoid problems with equation \ref{tauWindDipole} when $D_\text{p} \rightarrow 0$ in the limit $L_\text{p}D_\text{p}^2 \gg 4\dot{E}(1-D_\text{p})$ as discussed in \citet{harding99}.

Despite the magnetar-like outbursts of J1846--0258 and J1119--6127, the HBPs in table \ref{tableSys} appear to be mostly rotation-powered since $L_\text{x} < \dot{E}$. Therefore we also consider acceleration gap models for particle winds developed for rotation-powered radio pulsars \citep{windRS, windSpitkovsky, windPulsarDeath, windXuQiao}. A general feature of these models is the acceleration of charges through the potential above the NSs poles, such that the resulting spin-down is a sum of magnetodipole radiation and the energy that is carried away by the accelerated particle wind
\begin{equation}
\dot{E}_\text{cap} = \frac{2 \mu^2 \Omega^4}{3c^3} \left( \sin^2 \theta + 3 \kappa \frac{\Delta \phi}{\Delta \Phi} \cos^2\theta \right)  = \dot{E}_\text{d} \eta_\text{cap},
\label{EdotCap}
\end{equation}
where the particle density of the magnetosphere is a factor of $\kappa$ greater than the Goldreich-Julian charge density \citep{gjDensity}, $\Delta \phi$ is the potential within the acceleration region, and the maximum acceleration potential for a rotating dipole is $\Delta \Phi=\mu \Omega^2 / c^2$ \citep{windRS}. The particulars of the acceleration region determine $\Delta \phi$ , and a number of distinct examples can be found in the literature \citep{windXuQiao}. Despite the specifics, this family of models has a parametric form
\begin{equation}
\eta_\text{cap}= \sin^2 \theta + \xi \kappa B_\text{p}^{-u} \Omega^{-v} \cos^2 \theta
\label{etaCap}
\end{equation}
 where $\xi$, $u$ and $v$ are model dependent numerical constants. We allow for an offset of the rotation and magnetic poles by an angle $\theta$. The factor of $\cos^2 \theta$ is left out by some sources \citep{wind1846} without changing the physics by redefining $\kappa \rightarrow \kappa \cos^2 \theta$. For a review of acceleration gap models, see table $2$ in \citet{windPulsarDeath}. Let us also consider a time-dependent particle density in the magnetosphere, $\kappa \rightarrow \kappa(t)$. This gives the braking index
\begin{equation}
n_\text{cap}=3+\frac{\Omega}{\eta}\frac{\text{d}\eta}{\text{d}\Omega} + \frac{\kappa}{\eta} \frac{\text{d}\eta}{\text{d}\kappa}\frac{\Omega}{\dot{\Omega}} \frac{\dot{\kappa}}{\kappa}.
\label{nCapGap}
\end{equation}

Though the magnetar wind model looks distinct from the acceleration gap models, we will show here that it has the same parameterization as an acceleration gap model and produces similar results in terms of evolution. The similarity between equations \ref{etaCap} and \ref{etaWD} suggests we equate the role of the duty cycle in the magnetar wind model with the inclination angle $D_\text{p} \rightarrow \cos^2 \theta$ in equations \ref{etaWD} and \ref{etaCap}, giving a luminosity
\begin{equation}
L_p = \xi^2 \kappa^2 B_p^{-2(u-1)} \Omega^{-2(v-2)} \frac{R^6}{6c^3},
\label{LpRot}
\end{equation}
Thus the luminosity in the acceleration gap models depends on both the magnetic field and the rotational frequency of the NS. The time derivative of the luminosity is then
\begin{equation}
\dot{L_\text{p}} = 2 L_\text{p} \left[ \left(2-v\right) \frac{\dot{\Omega}}{\Omega} + \frac{\dot{\kappa}}{\kappa} \right]
\end{equation}
Substituting this form for the luminosity into the magnetar wind braking index (equation \ref{nWind}) reproduces the acceleration gap result from equation \ref{nCapGap}, evaluated with $\eta_\text{cap}$:
\begin{equation}
n = 3 + \frac{R^6}{6c^3} \frac{D_\text{p} }{I \dot{\Omega}} \xi \kappa B_p^{2-u} \Omega^{3-v} \left( v - \frac{\Omega}{\dot{\Omega}} \frac{\dot{\kappa}}{\kappa} \right) .
\label{nCap}
\end{equation}
The relations above show that the acceleration gap and magnetar wind model are closely related and produce the same basic evolutionary behavior despite the apparent distinctions between them. However, these models are not exactly degenerate because the minimum braking index for the gap models is $n<1$. We will explore the magnetar model and, for comparison with the literature, make use of the vacuum gap model with curvature radiation, labelled the VG(CR) model, which has $u=8/7$, $v=15/7$, and $\xi=496$ in units of $10^{12}$ G and $10^6$ cm.

\begin{figure}
\centerline{\includegraphics[scale=0.80, bb= 150 246 440 544, clip=true]{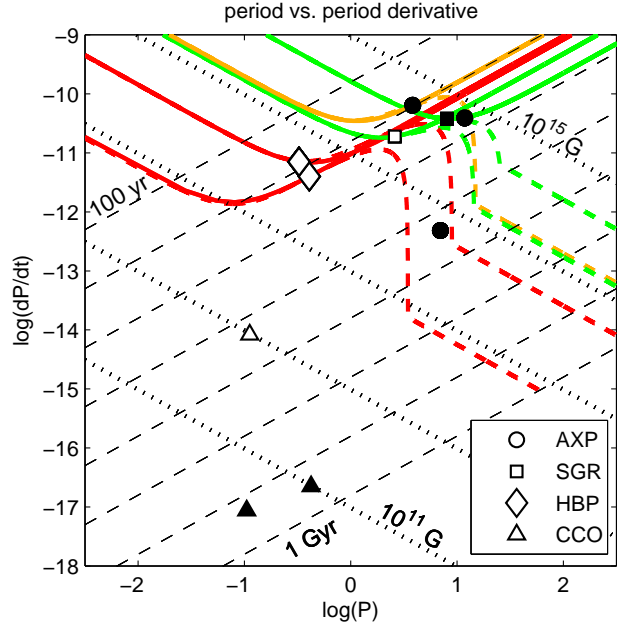}}
\caption{Period vs. period derivative for wind models applied to systems with confirmed or candidate wind nebulae. Solid lines show the magnetar wind model, and dashed lines show the analogous VG(CR) model, with the effect of pulsar death included. Red lines are the evolutionary trajectories of the HBPs, green lines the SGRs and orange for the AXP CXOU J171405.7--381031. All other details are as in figure \ref{figPPdotBEvolution}.}
\label{figPPdotWind}
\end{figure}

The dynamics of wind dominated systems can be significantly altered by including the effect of pulsar death \citep{windPulsarDeath}. As a NS loses energy, its rotational frequency decreases until it matches a critical `death period', where the maximum potential drop is insufficient to accelerate particles through the gap ($ \Delta \phi < \Delta \Phi$). At this stage of evolution the magnetospheric particle density is not replenished, and radio emission ceases. After the death period is reached the NS spins down solely by emission of magnetodipole radiation. Pulsar death introduces a factor $(1-\Omega_\text{death}/\Omega)$ to the second term of equation \ref{etaCap}, where the death period is
\begin{equation}
P_\text{death}=2.8 \left( \frac{B_\text{p}}{V_\text{n}} \right)^{\frac{1}{2}}
\end{equation}
with the polar magnetic field $B_\text{p}$ in terms of $10^{12}$ G, and $V_\text{n}$ the gap potential in terms of $10^{12}$ Volts \citep{windSpitkovsky, lowInclinationSGR}. In figure \ref{figPPdotWind} we plot the evolutionary trajectories given by the magnetar wind model as solid lines and show the acceleration gap models including the effect of pulsar death as dashed lines. In this case we have specified identical initial conditions to compare the evolution of each of the systems examined.

\begin{figure}
\centerline{\includegraphics[scale=0.8, bb= 209 219  386 573, clip=true]{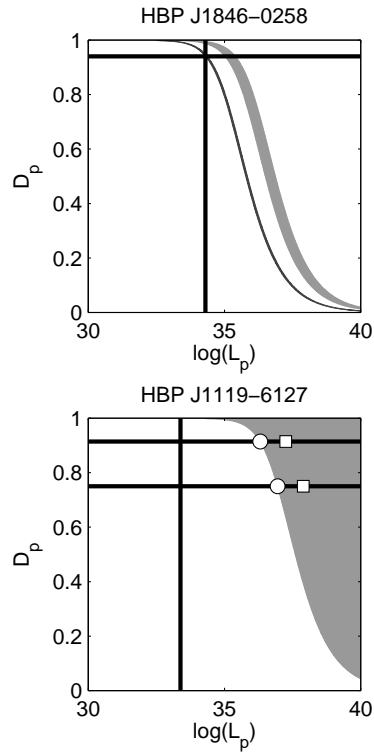}}
\caption{Dipole and wind age parameters for the HBPs J1846--0258 (top) and J1119--6127 (bottom). We plot the particle luminosity $L_p$ against the duty cycle $D_\text{p}$. In the upper panel, the pre- and post-outburst phases of J1846--0258 are shown in dark and light gray, respectively. The horizontal line indicates the duty cycle limit at which $L_\text{p}=L_\text{x}$ (vertical line). In the lower panel, parameters satisfying the SNR age are in the gray region. The parameters for four solutions with variable luminosity discussed in the text are plotted as dots (low SNR age) and squares (high SNR age) for an initial spin period of $3$ ms. The X-ray luminosity $L_{\text{x}}$ is marked with a vertical black line, and the limits on the inclination angle (duty cycle) are shown as black horizontal lines.}
\label{figWindPar}
\end{figure}

The luminosity and duty cycle parameters for the HBPs are plotted in figure \ref{figWindPar}. This figure shows the region of the $L_\text{p}-D_\text{p}$ parameter space that satisfies the age constraint of J1846--0258  in the top panel, and J1119--6127 in the bottom panel. The pre-outburst and post-outburst constraints are plotted as dark gray and light gray respectively. The changing braking index of J1846--0258 after the 2008 burst event requires a time-varying luminosity, but since no significant change in the X-ray pulse profile occurred any such change is strongly constrained \citep{archibald15}. We assume the derivative is negligible before and has a finite value after the burst, episodically varying about the mean on long timescales. The braking index is calculated by fitting the pre-outburst state without a derivative term, and then solving for the required instantaneous derivative $\dot{L}_\text{p}$ from equation \ref{nWind}. An upper limit on the duty cycle exists from the condition that the wind luminosity must equal or exceed the X-ray luminosity. Thus, for a minimum luminosity of $L_\text{p}=L_\text{x}=2 \times 10^{34}$ erg s$^{-1}$ with $D_{\text{p}} < 0.947$ ($\theta=14.08^\circ$) using an initial period of $3$ ms, giving an SNR age of $805$ years. This solution requires $\text{d}{L}_\text{p}/\text{dt}=2.82 \times 10^{24}$ erg s$^{-2}$, implying that even a relatively small luminosity growth rate after the glitch event has significant impact on the braking index. To compare with the results of previous studies that used the VG(CR) model, we fixed the inclination angle at $\theta = \pi / 4$ as used in \citet{wind1846}. At this inclination we find a polar magnetic field $\sim 1.25 \times 10^{14}$ G, density factor $\kappa$ between $56$ and $59$ growing at a rate $2.55 \times 10^{-9} \leq \dot{\kappa} \leq 4.63 \times 10^{-9}$. These values are similar to the results of \citet{wind1846} after adjusting for the extra factor of $1/2$, omitted by convention in that study, and using $n=2.19$ \citep{archibald15}. The duty cycle $D_\text{p}$ (or the corresponding inclination angle) strongly affects the solution, so knowledge of the system geometry is crucial to improve the results.

As mentioned in Section \ref{sec:B(t)}, the braking index for J1119--6127 with a constant luminosity produces a braking index that is smaller than observed ($n \sim 1.2$) when using the $4.2$ to $7.1$ kyr age range of G292.2--0.5, though the age is matched well. Conversely, a constant braking index with the contouring approach gives solutions up to a maximum age of $t_{\text{SNR}}\approx 1.76$~kyr. This result is similar to the prediction of the field growth model (Section \ref{sec:B(t)}). This problem vanishes when making use of a wind model with a variable luminosity. The region of parameter space with acceptable solutions for J1119--6127 is shown on the lower panel of figure \ref{figWindPar} as a light gray area and requires a decreasing luminosity $\dot{L_\text{p}}<0$ which gives a lower braking index in the recent past as suggested by \citet{harsha1}. The inclination angle of J1119--6127 has been determined to be in the range $17^{\circ} \leq \theta \leq 30^{\circ}$ \citep{welt11}, and the corresponding duty cycle is then between $0.91$ and $0.75$ respectively. These limits are plotted as horizontal lines in figure \ref{figWindPar}. To set limits on the parameters we consider an initial period of $3$ ms at the duty cycle extremes for the minimum and maximum SNR age, plotted as disks and squares respectively. This gives luminosity between $2.04 \times 10^{36}$ and $7.72 \times 10^{37}$ erg s$^{-1}$, with time-derivative between $-2.54 \times 10^{27}$ and $-6.61 \times 10^{25}$ erg s$^{-2}$.

\section{Discussion}
\label{sec:discussion}

We account for the X-ray luminosity of the AXPs using a decaying field, except for the AXP 1E~2259+586. For this system field decay was insufficient to explain the luminosity by a factor between 2 and 20. This discrepancy was found in previous work \citep{dall'osso} and led to the suggestion of an additional internal decaying toroidal field component.

The joint fits for the decaying field in Figure \ref{figTAge} also provide some interesting suggestions. The error bar on the age of the remnant G348.7+00.3, associated with AXP CXOU~J171405.7--381031 is large and crosses the line for which $\tau_{\text{PSR}}=\tau_{\text{SNR}}$. If CXOU~J171405.7--381031 is like the other AXPs then the NS characteristic age should appear older than the SNR and it has a braking index $n>3$. In this case the joint fit for field decay unites the AXPs as snapshots in the evolution of these objects. If the converse is true the remnant appears older than the AXP and the braking index is $n<3$, implying that field decay is the wrong explanation for this behavior, so some other mechanism must influence the behavior of the NS. These comments also apply to the SGRs, which do not have conclusive lower SNR age limits, and from this perspective it is not clear if these sources evolve by field growth or decay. The AXP CXOU~J171405.7--381031 and the SGRs 0526--66 and 1627--41 are similar in that they all have a $\tau_{\text{PSR}} < \tau_{\text{SNR}+}$, and there is no distinct age discrepancy. Since no lower age estimates for N49 or G337.3--0.1 exist, the SGR parameters cannot be tightly constrained in model fitting. In order to gain more insight on these systems it is critical to tighten the age estimates of the associated SNRs, and future measurements of the braking index (whenever possible) would conclusively settle this issue.

Both of the HBPs provide extremely interesting insights on energy-loss mechanisms. The HBP J1119--6127 shows a faint wind nebula that is very similar in characteristics (e.g., $L_\text{x}/\dot{E}$ and photon index) to wind nebulae surrounding rotation-powered pulsars \citep{samarHarsha1119}. Our study confirms these expectations and shows a changing particle luminosity is needed for this system. We find the required instantaneous luminosity derivative, but more detailed scenarios have been discussed in the literature \citep[see, for example, ][]{windbraking2013}.

Variability in the wind nebula around HBP J1846--0258 has been observed in reaction to outburst activity of the NS \citep{ng1846}. The wind emission mechanism can account for both pre- and post-burst states with the age and braking index constraints with a varying particle luminosity. \citet{archibald15} argue against the applicability of the wind braking mechanism in HBP J1846--0258, and favour an explanation based on changes in the structure of the magnetosphere. However, \citet{wind1846} have explained the difference in terms of a time-dependent wind density, which our work supports. \citet{wind1846} argue that the necessary wind density enhancement is on the order of $\sim 1 \%$, which provides a small change to the observed luminosity but is significant enough to explain the observed change in braking index. We also provide a link between the various wind models such as the magnetar wind model presented in \citet{harding99} and the various forms of acceleration gap models used in the literature \citep{windXuQiao}.

In general, the dependence of the Hall drift timescale on the magnetic field (equation \ref{hallTau}) determines the physics at work in describing NS evolution. \citet{hallRef2} has analyzed the properties of a sample of $118$ NSs, most of which are recycled millisecond PSRs with measured braking index and relatively low magnetic fields. Their Figure 2 shows the evolutionary properties of this sample of low-field NSs. A number of conclusions are drawn based on this population, namely that the objects transition from an initial state with braking indices $n<3$ at early times to $n>3$ from field decay on Ohmic timescales between $10$ and $100$ kyr. Moreover, \citet{hallRef2} also show the results of simulations which use all possible decay modes of the magnetic field including both Hall drift and Ohmic dissipation, along with an early period of accretion to bury the magnetic field. The evolutionary trajectories for submerged fields shown in figure \ref{figPPdotBEvolution} reproduce the qualitative behaviour from the simulations for a substantial amount of accreted mass $\approx 10^{-3} M_\odot$. This scenario produces a vertical trajectory in $P$ and $dP/dt$ at early times. However, simulations with a shallow submerged field under lower accreted mass ($\approx 10^{-5}$ to $10^{-4}$) nearly follow along lines of constant magnetic field, relevant for the HBPs. The simulations presented in \citet{hallRef2} predict that the braking indices of the HBPs will be mainly affected by field decay and should generally increase with time. \citet{hallRef2} also provide several examples of additional, potentially dominant, effects such as alignment of the magnetic field and magnetospheric variability. Since the environs surrounding the young high-field PSRs are complex and display wind nebulae \citep{samarHarsha1119, samarMWN, younes16}, short-timescale variations due to the emission of particle winds should also be included as an evolutionary effect significant for the variation of braking index and spin. The evolutionary tracks in figure \ref{figPPdotWind} for wind emission predict that the braking index should decrease steadily with time, however the variable nature of the wind luminosity needed to explain the post-outburst braking index of J1846--0258 argues against such an orderly evolution. Realistically both magnetothermal and particle wind effects must play important roles in the behaviour of the HBPs.

\section{Conclusions}
\label{conclusions}

We have analysed several mechanisms to produce the spin evolution of `anomalously' magnetized neutron stars by studying energy-loss mechanisms constraining the NS characteristic age and SNR age to agree. We also use the X-ray luminosity and braking index constraints when available. For constant values of the braking index we find a large ratio $P_0/P$ is often required to satisfy the age constraints, at odds with the prediction of the magnetar model. This leads to a discussion of emission mechanisms in which the braking index is time-dependent.

For the AXPs, which have $\tau_{\text{PSR}}>\tau_{\text{SNR}\pm}$, we favour the field decay scenario which predicts $n>3$, and we explored the properties of these solutions in terms of NS evolution. We performed a joint fit to the AXPs simultaneously and found a family of solutions containing those previously studied by \citet{nakano15}. The CCOs require exponential decay to explain their joint behavior, which is an unlikely decay mode for physical reasons. However, the population of CCOs can be neatly described in terms of field growth, which links apparently disparate classes of NS by evolution. This provides an evolutionary link between NSs in the HBP and CCO regions of the $P-\dot{P}$ phase diagram \citep[][RSH16]{ho15}. Despite the success of field growth in isolated NSs, the two HBPs that are associated with wind nebulae inside SNRs are best described in terms of a wind model \citep{harding99, windbraking2013, wind1846}. The wind model was able to account for the J1846--2058 pre- and post-outburst braking index, remnant age and X-ray luminosity but requires a variable particle luminosity. In our analysis we relate the magnetar wind and acceleration gap models and show they describe the same qualitative evolutionary trajectories through the $P-\dot{P}$ phase space, despite the unique physical assumptions included in each scenario.

Constraining the SNR ages will produce tighter constraints on the parameters for each of the models considered in this work. Finally, increasing the sample of pulsar-SNR secure associations and improving the age, distance estimates and braking indices are crucial to expand on this study.

\section{Acknowledgements}
This work was primarily supported by the Natural Sciences and Engineering Research Council of Canada (NSERC) through the Canada Research Chairs Program.
SSH also acknowledges support by an NSERC Discovery grant and the Canadian Space Agency.
This research made use of NASA's Astrophysics Data System, McGill's magnetars catalog and the U. of Manitoba's high-energy SNR catalogue (SNRcat).
We thank Harsha Kumar for discussions on magnetar SNRs and Gilles Ferrand for contributions to SNRcat. We also extend thanks to the anonymous referee for providing a valuable review that improved the overall quality of the text.

\clearpage

\label{lastpage}

\setcounter{table}{0}
\begin{landscape}
\begin{table}
\begin{center}

\begin{tabular}{|l|l|l|l|l|l|l|l|l|l|l|}
\multicolumn{11}{c}{Observed properties of AXPs, SGRs, HBPs, and CCOs associated with SNRs} \\
\hline

PSR & $P$ & $\dot{P}$ & $n$   & $L_x$ & $\dot{E}$ & $\tau_{PSR}$ & $B_0$ & SNR & $\tau_{SNR-}$ & $\tau_{SNR+}$ \\
    & s  &$10^{-11}ss^{-1}$ & &$10^{33}$ erg s$^{-1}$&$10^{33}$ erg s$^{-1}$ &kyr & $10^{14}$ G &  & kyr & kyr \\ \hline
\hline

AXP~1E~1841--045 & $11.79$ & $4.09$ & $$ & $184$ & $0.99$ & 4.75 & 7.05 &  G27.4+0.0 (Kes~73)  & $0.75$  & $2.1 $ [1]  \\ \hline
AXP~1E~2259+586 &  $6.98$  & $4.8e-2$ & $$ & $17$ &  $5.3e-2$ & 228.317 & 0.59 & G109.1-01.0  (CTB~109)& $10$ & $16$ [2] \\ \hline
CXOU~J171405.7--381031 & $3.83$  & $6.40$ & $$ & $56$ & $45.14$ & 0.95 & 5.0 & G348.7+00.3 & $0.35$ & $3.15$ [3] \\ \hline
\hline

SGR~0526--66 & $8.05$ & $3.8$ & $$ & $189$ & $2.87$ & $3.358$ & $5.60$ & N49 & $-$ & $4.8$ [4] \\ \hline
SGR~1627--41 & $2.59$ & $1.9$ & $$ & $3.6$ & $33.9$ & $2.164$ & $2.25$ & G337.3--0.1 & $-$ & $5.0$ [5]  \\ \hline
\hline

PSR~J1119--6127   & $0.408$ & $0.40$ & $2.684\pm 0.002$ {[11]}& $2.4$     & $2325$  & 1.616   & 0.41   & G292.2--0.5 & $4.2$ & $7.1$ [6] \\ \hline
PSR~J1846--0258 A & $0.325$ & $0.71$ & $2.64 \pm 0.01$ {[12]} & $20.0$    & $8161$  & $0.726$ & $0.485$ & G029.7--0.3 (Kes~75)& $0.430$ & $4.3$ [7] \\
PSR~J1846--0258 B & $0.327$ & $0.71$ & $2.16 \pm 0.13$ {[13]} & $$            & $8056$  & $0.728$ & $0.49$ &                     &         & \\ \hline
\hline

RX~J0822.0--4300 & $0.112$ & $8.3e-4$ & $$ & $5.6$ & $233$ & $213.799$ & $9.80e-4$ & G260.4--3.4 (Puppis A) & $3.70$ & $5.20$ [8] \\ \hline
1E~1207.4--5209 & $0.424$ & $2.23e-6$  & $$ & $2.50$ & $1.16e-2$ & $3.01e5$ &$9.85e-4$ & G296.5 +10.0 (PKS 1209--51/52) & $2.0$ & $20.0$ [9] \\ \hline
CXOU~J185238.6+004020 & $0.105$ & $8.68e-7$  & $$ & $5.30$ & $0.296$  & $1.92e5$ & 3.06e-4 & G033.6+00.1 (Kes 79) & $5.4$ & $7.5$ [10] \\ \hline

\end{tabular}
\caption{For a given PSR, $P$ is the period, $\dot{P}$ the period derivative, and $L_x$ the X-ray luminosity in the range $2$ to $10$ keV from the McGill magnetar catalogue (\protect\url{http://www.physics.mcgill.ca/~pulsar/magnetar/main.html}). The equatorial magnetic field strength in flat space-time is $B_0$. The NS characteristic age is $\tau_{PSR}$ and the SNR age limits are $\tau_{SNR-}$ and $\tau_{SNR+}$. The SNR ages have been compiled in the U. of Manitoba's SNR Catalogue (SNRcat, \citet{SNRCat},\protect\url{http://www.physics.umanitoba.ca/snr/SNRcat/}). References to SNR ages in this table are [1]: \citet{1841Age}, [2]: \citet{nakano15}, [3]: \citet{1714Age2009}, [4]: \citet{0526Age}, [5]: \citet{1627Age}, [6]: \citet{harsha1}, [7]: \citet{1846Age}, [8]: \citet{0822Age}, [9]: \citet{1207Age}, [10]: \citet{1852Age}. References to the braking indices included here are [11]: \citet{welt11}, [12]: \citet{livingstone06}, [13]: \citet{livingstone11}.}
\label{tableSys}
\end{center}
\end{table}
\end{landscape}

\appendix
\section{Excluded Systems}
\label{appA}
There are a number of systems that are absent from table \ref{tableSys} which are not secure NS-SNR associations, or have unreliable SNR age estimates. In this section we address the reasoning behind our selections.

The SGR 0501+4516 and SNR HB9 \citep[G160.9+2.6;][]{SGR0501} have an uncertain association. The pulsar is located just outside the SNR shell and so would require a high projected space velocity of $\sim1700$--$4300$ km s$^{-1}$ for a distance to the SNR of $1.5$ kpc and an age of $8$--$20$~kyr \citep{SGR05012008}, which argues against the association.

SGR 1806--20 and the object G10.0--0.3  \citep{diskMarsden} have been claimed to be associated with one another, but G10.0--0.3 is no longer considered a SNR and is now thought to be a radio nebula powered by a stellar wind \citep[cf.][]{greensCatalogue}.

The SGR 1E~1547.0--5408 is likely associated with G327.24--0.13, but the SNR age is highly uncertain. \citet{1E1547} assume an SNR age $<1.4$ kyr based on the association with the NS.

We also exclude SGR 1900+14 from our sample, which is in a complex region of the sky containing many SNRs around its current position \citep{SGR1900complex}. In fact, SGR~1900+14 is believed to be associated with a cluster of massive stars \citep{SGR1900cluster}. Furthermore, SGR~1900+14 appears to be separate from the SNR G42.8+0.6 \citep{marsden1999, thompson2000}, and would require a large recoil velocity if related. Since the SNR association of SGR 1900+14 is ambiguous we also exclude it from this study.

The interesting SGR Swift J1834.9--0846 shows strong evidence of interaction with its environment in the form of a wind nebula \citep{younes16, granot16}, and is associated with G023.3--00.3 \citep[W41;][]{1834Age2007}, though the age of the remnant is not well constrained and may vary between $60$ to $200$ kyr.

The PSR J1622--4950 is a transient magnetar, and is unlike any of the other sources we consider. This object is particularly interesting because it is the first to be discovered via its radio emission \citep{Levin2010} and its X-ray luminosity can be accounted for by its rotational energy loss (i.e. $L_x<\dot{E}$). It is proposed to be associated with the SNR candidate G333.9+0.0, but -- aside from the fact that the SNR's nature is yet to be confirmed and the association is considered unlikely \citep{PSR1622} -- the SNR age is also not reliable. \citet{PSR1622} provide an upper limit only on the Sedov age of G333.9+0.0 to be $\tau_{SNR+}=6$ kyr, similar to the characteristic age of PSR J1622--4950, which is $\tau_{PSR}=4$ kyr. For these reasons we have also excluded PSR J1622--4950 from our sample.

Finally, we also exclude HBP J1734--3333, since the hypothesized association with the remnant G354.8--0.8 \citep{man} is still controversial. \citet{hoAn2012} cite the lower limit of the SNR age only with $t_{SNR} > 1.3$ kyr, based on the pulsar's distance from the remnant and its measured speed to obtain a lower limit on the SNR age.

\end{document}